\documentclass[prl,twocolumn,reprint,amsmath,letterpaper]{revtex4-1}
\usepackage{graphicx}
\usepackage{color}
\usepackage{amssymb}
\begin{document}

\title{Distinct contiguous versus separated triplet-pair multiexcitons in an intramolecular singlet fission chromophore}
\author{R. Chesler$^1$, P. Bhattacharyya$^2$, A. Shukla$^3$ and S. Mazumdar$^1$}
\affiliation{$^1$Department of Physics, University of Arizona Tucson, AZ 85721, USA}
\affiliation{$^2$Institute for Theoretical Solid State Physics, Leibniz IFW Dresden, Helmholtzstraße 20, 01069 Dresden, Germany}
\affiliation{$^3$Department of Physics, Indian Institute of Technology Bombay, Powai, Mumbai 400076, India}

\begin{abstract}
We show from many-body quantum mechanical calculations that there occur structurally distinct triplet-pair eigenstates
in the intramolecular singlet fission (iSF) compound pentacene-tetracene-pentacene. Triplet excitons occupy neigboring pentacene and tetracene monomers
in the higher energy doubly degenerate triplet-triplet multiexcitons, and terminal pentacene chromophores in the lower energy multiexciton. 
The lowest energy multiexciton is reached 
by ultrafast triplet migration within the triplet-triplet manifold, a result
with profound implication for the design of superior iSF compounds.
\vskip 0.5pc
\end{abstract}

\maketitle

Electron correlation effects on the optoelectronic properties of carbon-based $\pi$-conjugated 
systems have been of continuous and intense interest \cite{Baeriswyl92a,Soos94a,Ramasesha00a,Barford05a}. 
One consequence of strong $\pi$-electron correlation is the occurrence of the two electron-two hole (2e-2h) bound spin triplet-pair state 
$^m$(T$_1$T$_1$) energetically 
below the one electron-one hole (1e-1h) optical spin-singlet state in long linear 
conjugated polyenes \cite{Hudson82a} and polyacenes \cite{Raghu02a,Sanders20a}
(here and in what follows T$_1$ is the lowest triplet exciton, and $m$ the overall spin multiplicity).  
This energy ordering is of direct relevance to singlet fission (SF), a spin-allowed photophysical process involving the internal 
conversion of the optically accessible spin-singlet 
exciton to the optically dark $^1$(T$_1$T$_1$) state
\cite{Smith13a,Lee13a,Rao17a,Xia17a,Casanova18a,Felter19a}. 
The possibility of circumventing the Shockley-Queisser limit to photoconductivity \cite{Shockley61a} in organic solar cells
has been the driver of the field.
In systems with small triplet-triplet binding energy E$_b$, defined as 2$\times$E(T$_1$) - E($^1$(T$_1$T$_1$)), where E(T$_1$)
and E($^1$(T$_1$T$_1$)) are the energies of the lowest free triplet and the triplet-triplet multiexciton, 
$^m$(T$_1$T$_1$) will undergo dissociation to two free triplets T$_1$. In principle, each triplet   
can donate an electron to an acceptor molecule in a donor-acceptor heterostructure, thereby doubling the photoconductivity \cite{Smith13a,Lee13a,Rao17a,Xia17a,Casanova18a,Felter19a}. 

Until recently research on SF was largely limited to {\it inter}molecular SF (xSF), in which the triplet excitons constituting the $^1$(T$_1$T$_1$)
occupy chromophore monomers that are not bonded covalently \cite{Smith13a,Lee13a,Rao17a,Xia17a,Casanova18a,Felter19a}. SF being a multichromophore process requires moderate to strong ``through space'' coupling between the chromophore monomers
in this case,
which in turn places stringent morphological requirements for the chromophore composite, thereby limiting the classes of materials
available for xSF. The focus of SF research consequently shifted to {\it intra}molecular SF (iSF), in which the chromophore monomers are either directly
covalently bonded or are linked covalently via bridge molecules \cite{Sanders15a,Zirzlmeier15a,Lukman15a,Sakuma16a,Sanders16a,Korovina16a,Korovina18a,Zirzlmeier16a,Sun16a,Basel17a,Margulies16a,Pun19a,Krishnapriya19a,Korovina18c,Hetzer18a,Chen18a,Parenti20a,Wang22a,Purdy23a,Majumder23a}.
Based on ultrafast spectroscopic measurements that found excited state absorptions (ESAs) at triplet absorption energies of the terminal chromophore molecules
in the visible wavelength region
it was initially believed that iSF led to rapid generation of free triplets. Following
theoretical research that demonstrated that the bound $^1$(T$_1$T$_1$) had {\it additional} ESA in the IR over and above ESA
in the visible \cite{Khan17b,Khan18a}, and experimental observations of this additional ESA \cite{Trinh17a,Miyata19a}
it is now accepted that while iSF may lead to rapid $^1$(T$_1$T$_1$) generation, quantitative dissociation to free triplets is rare.
 
Successful implementation of iSF in organic photovoltaics will require fast $^1$(T$_1$T$_1$) generation 
as well as its fast dissociation to free triplets, overcoming competing photophysical processes that include triplet recombination.
This requirement poses serious theoretical and experimental challenges, as fast $^1$(T$_1$T$_1$) generation necessarily requires strong effective electronic coupling
between the chromophore components occupied by the individual triplet excitations of the $^1$(T$_1$T$_1$)
wavefunction \cite{Pensack18a,Pun19a,Masoomi-Godarzi20a}. 
The latter in turn leads to strong E$_b$, which slows down and even prevents triplet dissociation.
Recent iSF research has therefore focused on the search for appropriate chromophore-bridge molecule combination
that can solve the riddle of fast $^1$(T$_1$T$_1$) generation yet small E$_b$. From a theoretical perspective
this requires, (i) precise understanding of $^1$(T$_1$T$_1$) wavefunction, (ii) the dependence of the triplet-triplet entanglement 
on the structure of the bridge molecule, (iii) the mechanism of $^1$(T$_1$T$_1$) formation, and (iv) the mechanism of spin dephasing. 
Simply increasing the length of the bridge molecule is not an optimal solution; even as that reduces E$_b$ it also slows $^1$(T$_1$T$_1$) generation \cite{Parenti20a}. 
The $^1$(T$_1$T$_1$) has therefore been the focus of several recent reviews \cite{Kim18a,Musser19a}. 

An ingenious advance in the design of iSF compounds with ultrafast $^1$(T$_1$T$_1$) generation (in few ps) yet
long lifetime (hundreds of ns) was achieved recently by Pun {\it et al.} \cite{Pun19a}. The authors synthesized and tested a series of
iSF compounds {\bf P-Tn-P}, where {\bf P}, {\bf T} and {\bf n} refer to terminal pentacene chromophores, linker tetracene molecules
and the number of linker molecules, respectively (see Fig.~1). 
According to the authors, initial photoexcitation at pentacene singlet exciton energy creates a localized excitation on a terminal {\bf P} monomer,
which undergoes rapid conversion to a contiguous triplet-triplet pair $^1$(T$_{1[P]}$T$_{1[T]})$, where T$_{1[P]}$ and T$_{1[T]}$ refer to lowest
triplets on {\bf P} and {\bf T}, respectively. 
The difference in the triplet energies of {\bf T} and {\bf P} monomers ($\sim 0.3$ eV) then 
drives triplet migration and transition from contiguous triplets 
to the lowest energy triplet-pair $^1$(T$_{1[P]}$T$_{1[P]}$) in which the triplets occupy only the terminal pentacenes.
This conclusion was reached from observation of
transient ESA in the IR due to $^1$(T$_{1[P]}$T$_{1[T]}$) almost immediately after photoexcitation, and the rapid disappearance of this ESA (in 3.0 ps and 5.3 ps in $n=2$ and $n=3$, respectively) 
followed by $^1$(T$_{1[P]}$T$_{1[P]})$ generation \cite{Pun19a}.
The physical separation between the triplets and uphill recombination are behind 
the long $^1$(T$_{1[P]}$T$_{1[P]}$) lifetime \cite{Pun19a}.

\begin{figure}[b]
  \centerline{\resizebox{3.0in}{!}{\includegraphics{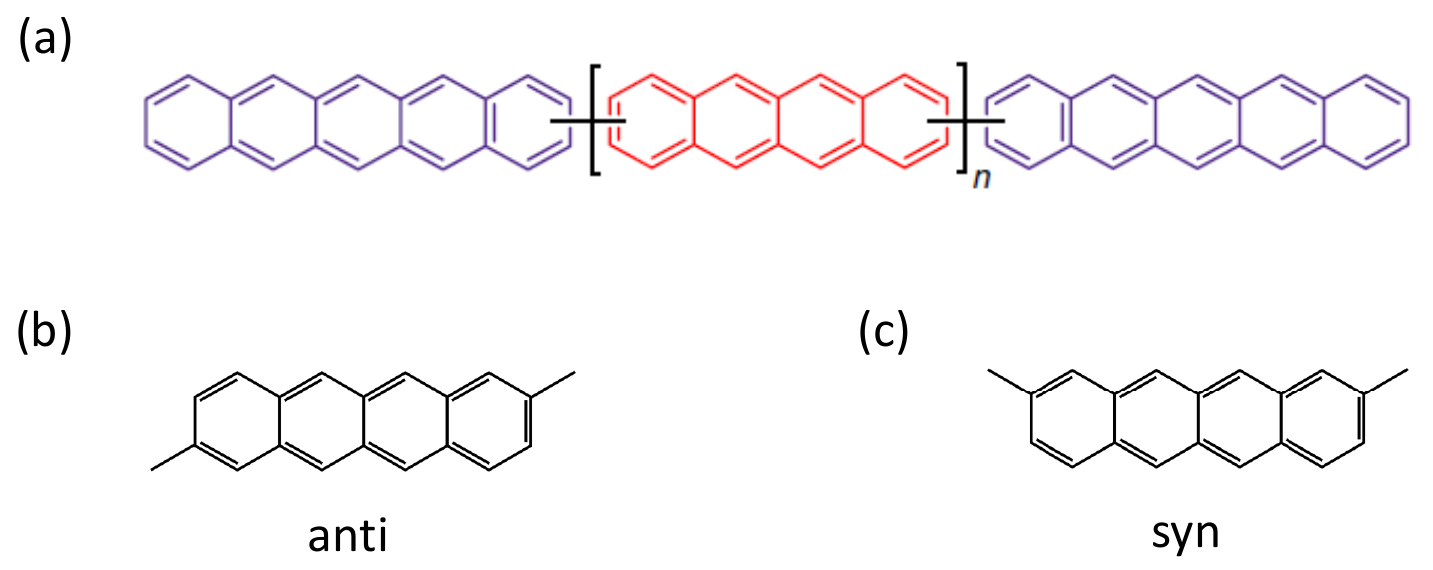}}}
  \caption{(Color online) (a) Schematic structure of the {\bf P-Tn-P} compounds investigated experimentally \cite{Pun19a}. (b) {\it Anti-} versus (c)
{\it syn-} linkages investigated here for the {\bf n=1} molecule.}
\end{figure}

The experimental results are exciting, as they for the first time suggest 
that triplet separation can occur by {\it downhill} triplet migration.
Theoretical evidence for distinct phase-coherent high energy $^1$(T$_{1[P]}$T$_{1[T]}$) and low energy $^1$(T$_{1[P]}$T$_{1[P]}$)
that would be required for the proposed triplet migration scheme \cite{Pun19a}, 
as opposed to quantum mechanical superpositions, currently does not exist. 
While weakly phase-coherent physically separated triplet-triplet  $^1$(T$_1 \cdots$ T$_1$) has also been
suggested in the context of xSF \cite{Scholes15a,Pensack16a,Lee18a,Miyata19a,Musser19a,Taffet20a,Hudson22a}, there the weakly bound multiexciton 
is reached from the strongly bound $^1$(T$_1$T$_1$) in an {\it uphill} process, overcoming the binding energy of the nearest neighbor triplet-pair. 
Convincing theoretical evidence for such a $^1$(T$_1 \cdots$ T$_1$) state in xSF is lacking. 
Furthermore, in structurally related compounds {\bf P}-$\beta$-{\bf P},
$\beta$ = benzene ({\bf B}), naphthalene ({\bf N}) and anthracene ({\bf A})
triplets in the lowest triplet-triplet $^1$(T$_1$T$_1$) do occupy the terminal {\bf P} chromophores only, but the $^1$(T$_1$T$_1$)
in these are reached via virtual charge-transfer (CT) excitation between the terminal pentacene chromophore themselves,
effectively bypassing the bridge molecules \cite{Parenti22a}. Taken together, these observations suggest that
the mechanism of iSF in these closely related compounds changes with the length of the bridge molecule. Clear
understanding of the roles of interchromophore CT versus triplet migration, as functions of electron correlation and bridge molecule 
length will be essential to establish triplet migration as the
dominant pathway to the lowest triplet-triplet in {\bf P-Tn-P}.

In what follows we present the results of detailed quantum many-body 
calculations of the excited state electronic structures of the {\bf n}=1 compound {\bf P-T-P}, focusing not only on the triplet-triplet states, but also
on the mechanism of their generation. 
Our calculations are based on the $\pi$-electron only Pariser-Parr-Pople (PPP) Hamiltonian \cite{Pariser53a,Pople53a}, which has 
been widely used to describe $\pi$-conjugated carbon(C)-based systems \cite{Baeriswyl92a,Soos94a,Ramasesha00a,Barford05a},
\begin{multline}
 H=\sum_{\langle ij \rangle,\sigma}t_{ij}(c_{i\sigma}^{\dagger}c_{j\sigma}+c_{j\sigma}^\dagger c_{i\sigma}) + U\sum_{i} n_{i\uparrow}n_{i\downarrow} \\ 
   +\sum_{i<j} V_{ij} (n_i-1)(n_j-1)
\label{PPP_Ham}
\end{multline}
Here $c^{\dagger}_{i\sigma}$ creates an electron with spin $\sigma$ on the $p_z$ orbital of C-atom $i$, $n_{i\sigma} = c^{\dagger}_{i\sigma} c_{i\sigma}$ is the number of electrons with spin $\sigma$ on atom $i$, and $n_i=\sum_{\sigma}n_{i\sigma}$ is the total number of electrons on the atom. 
$t_{ij}$ are nearest neighbor electron hopping integrals,
$U$ the Coulomb repulsion between two electrons occupying the $p_z$ orbital of the same C-atom,
and $V_{ij}$ is long range Coulomb interaction.
The parameters are taken from our previous applications of the
model to acene monomers and dimers {\cite{Khan17b,Khan18a,Parenti22a}. 
We have chosen peripheral and internal bond lengths for the acene monomers to be (1.40 $\mathring{\textrm{A}}$) and (1.46 $\mathring{\textrm{A}}$), respectively, and the corresponding $t_{ij}$ as $-$2.4 and $-$2.2 eV, respectively \cite{Khan17b,Khan18a}. 
We have assumed the molecules to be planar for 
simplicity, with the interunit
bond length 1.46 $\mathring{\textrm{A}}$ and the corresponding hopping integral $-2.2$ eV, respectively. Monomer
rotation effect can be taken into consideration by reducing the interunit $t_{ij}$ by a multiplicative factor of $cos \theta$, where $\theta$ is the dihedral angle \cite{Ramasesha90a}.
Explicit calculations have confirmed that physical conclusions are not altered substantively by ignoring rotation effects \cite{Khan17b}. 
We use the 
screened Ohno parameterization for the long range Coulomb repulsion, $V_{ij}=U/\kappa\sqrt{1+0.6117 R_{ij}^2}$,
where $R_{ij}$ is the distance in $\mathring{\textrm{A}}$ between C-atoms $i$ and $j$ and $\kappa$ is an effective dielectric constant \cite{Chandross97a}.
The Coulomb parameter for C-atoms $U$ (7.7 eV) and the dielectric constant $\kappa$ (1.3) are chosen based on fitting monomer energetics
\cite{Khan17b,Khan18a}.

Our calculations are done within the diagrammatic molecular exciton basis to obtain physical pictorial 
descriptions of eigenstates \cite{Khan17b,Khan18a,Parenti22a}.
Accurate determinations of energy orderings and correlated wavefunction of the $^1$(T$_1$T$_1$) excited states demand
inclusion of high order correlation effects. We use the multiple reference singles and
doubles configuration interaction (MRSDCI) procedure that incorporates CI with 
dominant me-mh excited
configurations (m=1-4) \cite{Tavan87a,Aryanpour15a}. The MRSDCI calculations are done over exciton basis active spaces of 22-26
localized molecular orbitals (8-10 MOs for each {\bf P} monomer and 6-8 per {\bf T} monomer,
see Supplemental Material (SM), S.1 \cite{SM}. 
Energy convergence required MRSDCI matrices of dimensions several times 10$^6$.
\begin{figure}[b]
  \centerline{\resizebox{3.5in}{!}{\includegraphics{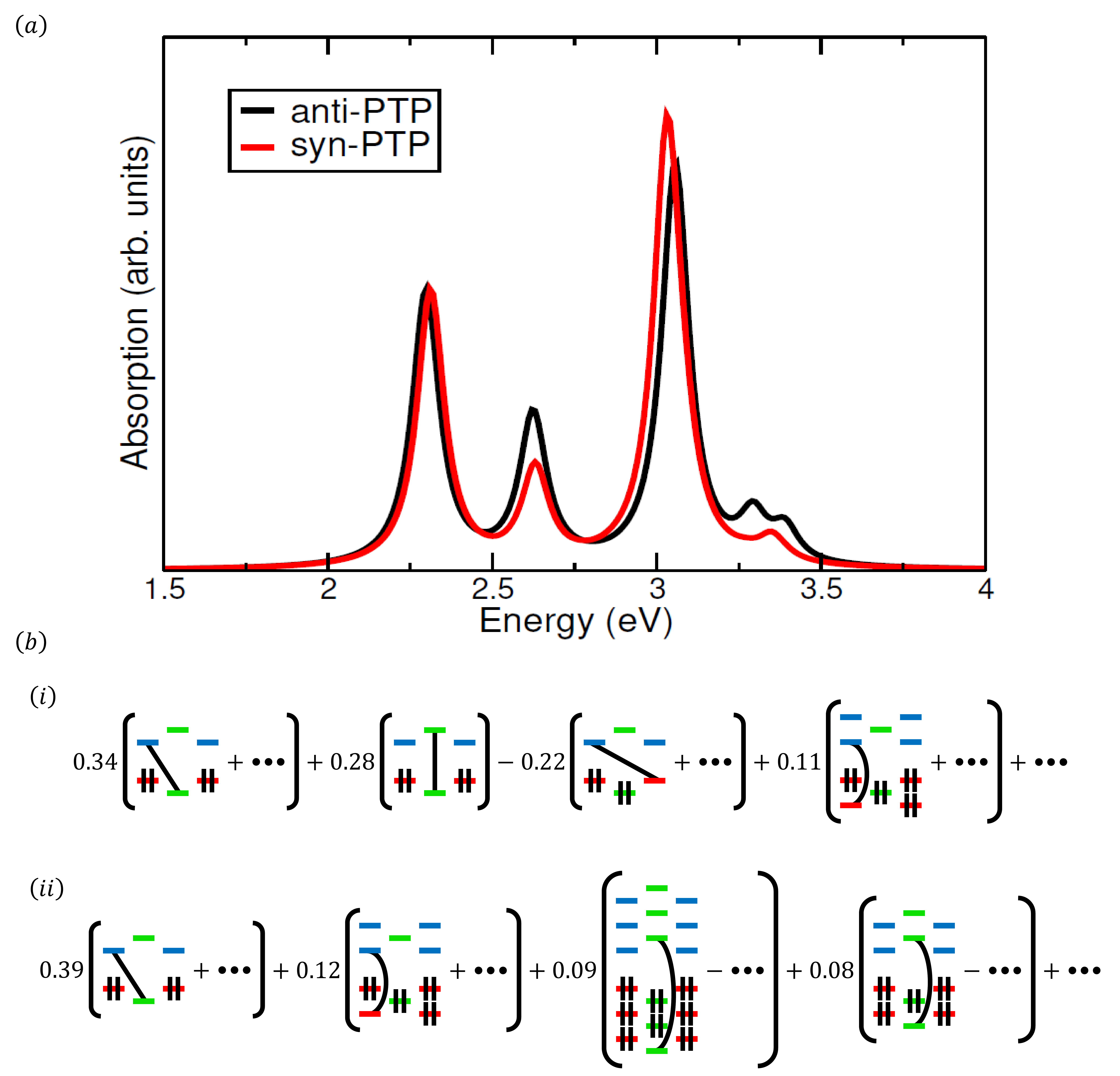}}}
  \caption{(Color online) (a) Ground state absorption spectra of {\it anti} and {\it syn}-{\bf P-T-P}. The two lower energy bands correspond to {\bf P}
and {\bf T} monomer absorptions, the highest energy band is due to intermonomer CT.
(b) Normalized exciton basis wavefunctions that contribute most strongly to the CT absorption band in (i) {\it anti-} and (ii) {\it syn-}{\bf P-T-P}. The horizontal bars correspond to {\bf P}
and {\bf T} HOMO and LUMO (see S.1 \cite{SM}). Electron occupancies of MOs are indicated. Straight lines connecting MOs are spin-singlet bonds. Ellipses correspond to 
additional configurations related by mirror-plane and charge-conjugation symmetries. CT excitation
between terminal {\bf P} monomers is much weaker than that
between nearest neighbor {\bf P} and {\bf T} in {\it anti-}{\bf P-T-P}, and is even weaker in the {\it syn-} compound. See also 
S.2 \cite{SM}}.  
\end{figure}

In Fig.~2(a) we have shown the calculated ground state optical absorption spectra for {\it anti}- and {\it syn}-{\bf P-T-P}.
The absorption spectra of Fig.~2(a) are 
different from those for {\bf P-B-P}, {\bf P-N-P} and {\bf P-A-P} in two important ways.
First, the CT absorption in {\bf P-T-P} occurs at energy higher than the localized monomeric absorptions of both the terminal pentacene and the linker tetracene, 
in contrast to {\bf P-B-P, P-N-P and P-A-P} where CT absorption is either to a state that occurs below the 
localized bridge molecule exciton ({\bf P-B-P} and {\bf P-N-P}) or to a state that exhibits significant configuration mixing with the bridge exciton ({\bf P-A-P}). The 
latter is the 
the criterion for bridge resonance \cite{Parenti22a} within correlated-electron theory.
Second, the nearly equal strengths of the CT absorptions for {\it anti}- and {\it syn}-{\bf P-T-P} is also in contrast to that in $\beta$ = {\bf B}, {\bf N} and {\bf A}, 
where the CT absorption occurs only for {\it anti} connectivity in {\bf P-B-P} and {\bf P-N-P}, and is 
stronger for {\it anti}-{\bf P-A-P} than in {\it syn}-{\bf P-A-P} \cite{Parenti22a}. 
The difference between {\it anti} versus {\it syn} connectivities there is ascribed to the tendency to electron-correlation driven antiferromagnetic spin-coupling 
between electrons on nearest neighbor C-atoms, which promotes selective stronger direct CT between the terminal pentacenes for {\it anti} connectivity.
The exciton-basis allows precise
characterizations of the final states of the absorption bands as excitations localized on the monomer molecules versus CT.  
Detailed examinations of the exciton basis CT eigenstates indicated one-to-one correspondence between iSF rate and the (a) strength of the dipole coupling between the
ground state and the CT eigenstate and (b) strong contribution to the CT eigenstate by the configuration with direct CT coupling between the terminal 
{\bf P} chromophores.
The constructive (destructive) quantum interference leading to strong (vanishing to weaker) CT absorption in {\it anti} ({\it syn}) $\beta$ = {\bf B, N, A} 
compounds is direct evidence for CT-mediated iSF \cite{Parenti22a}. 

We show in Figs.~2(b)(i) and (ii) the most
dominant exciton basis contributions to the CT eigenstates with the largest dipole couplings to the ground states in {\it anti}- and {\it syn}-{\bf P-T-P}. In strong
contrast to {\bf P-B-P}, {\bf P-N-P} and {\bf P-A-P}, the relative weight of the configuration with direct CT coupling between the terminal {\bf P} chromophores is 
noticeably smaller than that of configurations with CT between {\bf P} and {\bf T} (see (i)), or is vanishingly small (see (ii)).
Additional CT states also contribute to the very strong CT band in Fig.~2(a).
The energies, dipole couplings to the ground states and the dominant exciton basis contributions to all these states are
shown in supplementary Table I \cite{SM}. 
The relative weights of configurations with direct CT between the {\bf P} monomers are 
smaller than that of configurations with CT between {\bf P} and {\bf T}, or are vanishingly small in every case.
Dominant CT in all these wavefunctions is between
nearest neighbor {\bf P} and {\bf T} monomers, as in PT dimer \cite{Khan18a}, 
suggesting already that CT-mediation here can generate contiguous triplet-triplet $^1(T_{1[P]}T_{1[T]})$ but not 
$^1$(T$_{1[P]}$T$_{1[P]}$) with distant triplets. 

\begin{figure}[b]
  \centerline{\resizebox{3.5in}{!}{\includegraphics{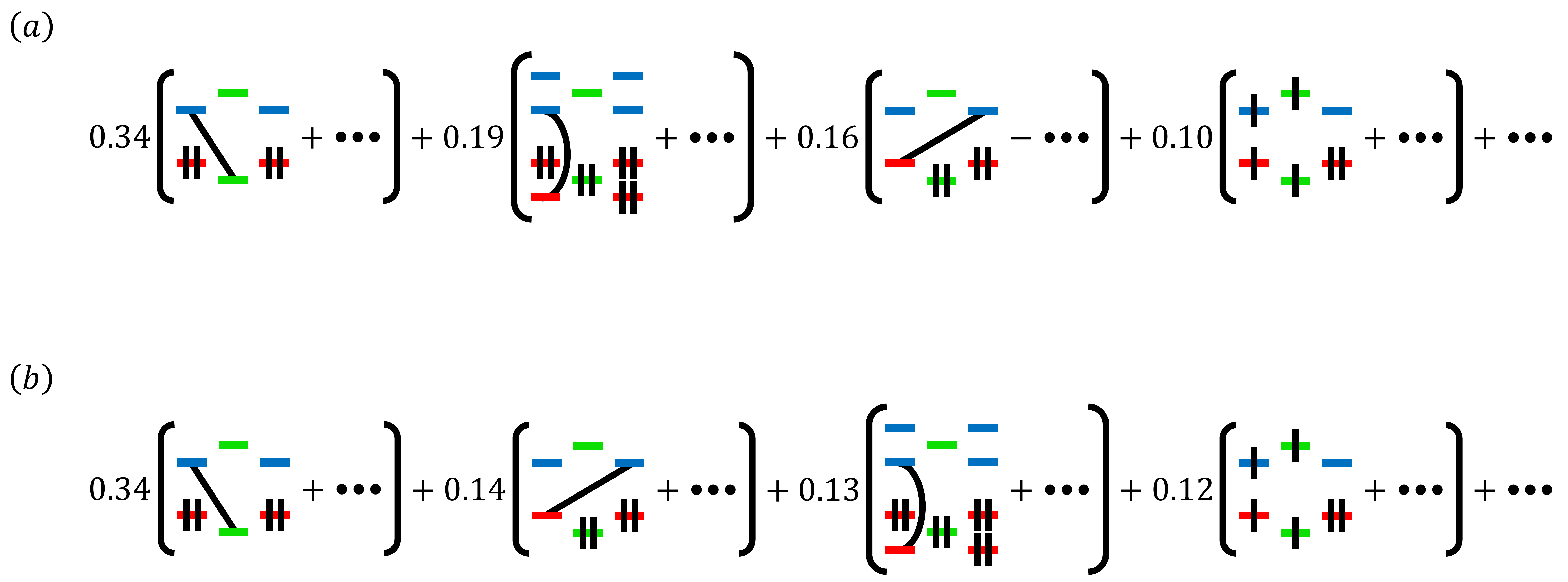}}}
  \caption{(Color online) 
Normalized wavefunctions of CT eigenstates to which ESAs from the optical spin singlet state are the strongest, for (a) 
{\it anti} and (b) {\it syn-}connectivity.}
\end{figure}

\begin{figure}[b]
  \centerline{\resizebox{3.2in}{!}{\includegraphics{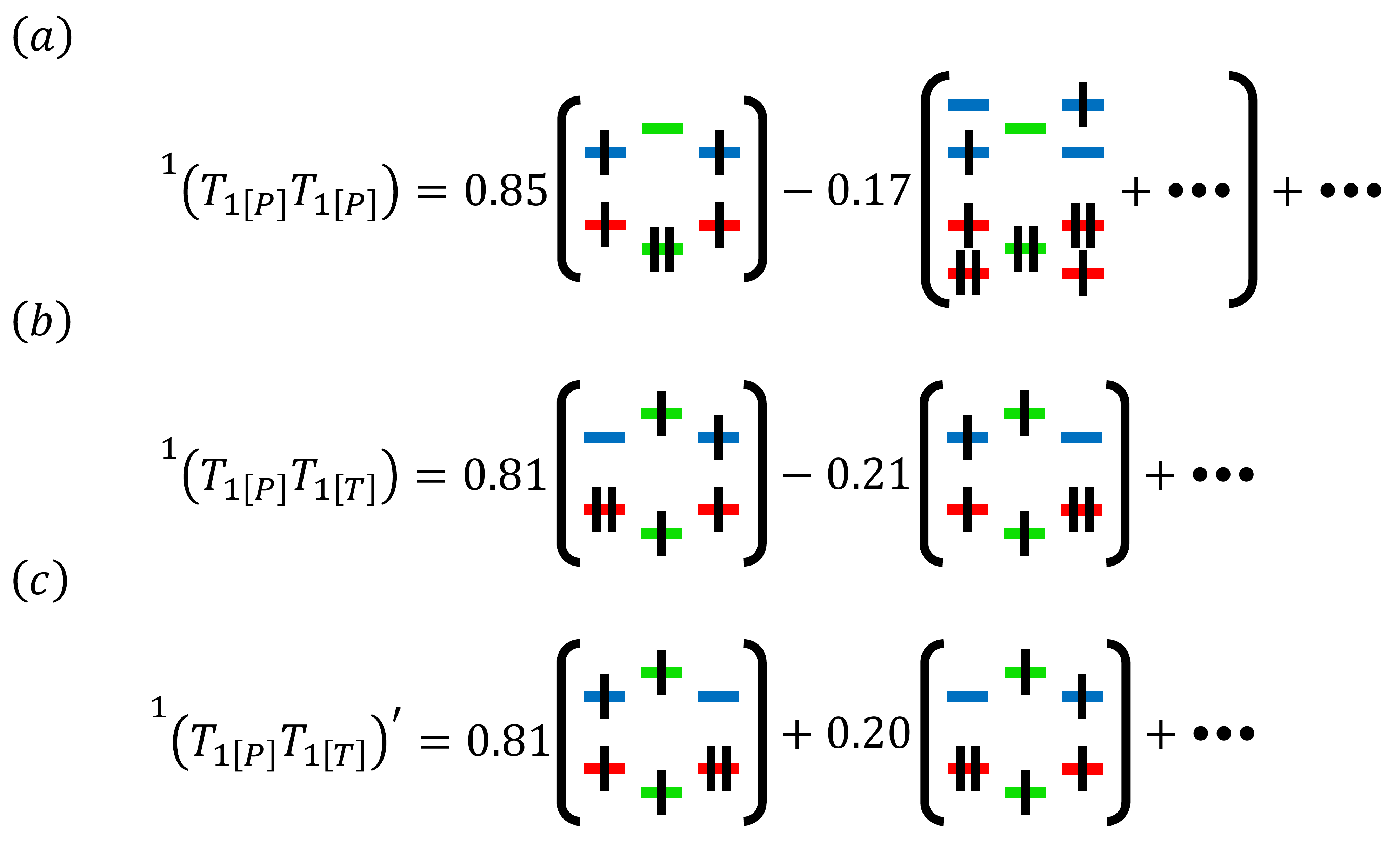}}}
  \caption{(Color online) Normalized wavefunctions of (a) $^1$[T$_{[1P]}$T$_{[1P]}$], (b) and (c) $^1$[T$_{[1P]}$T$_{[1T]}]$ states in {\it anti}-{\bf {\bf P-T-P}}. Intramonomer excitations are spin triplet, with the overall spin zero.
The corresponding wavefunctions for {\it syn}-{\bf {\bf P-T-P}} are nearly identical and are not shown.}
\end{figure}

CT wavefunctions of symmetric iSF molecules occur as nearly degenerate pairs with even and odd parity symmetries. Ground state absorption is to the
odd parity CT eigenstate and provides only indirect information about CT-mediated SF.
The CT process directly relevant for SF is the virtual excitation between the optical singlet and the even parity CT excitation 
\cite{Smith13a,Berkelbach13b,Yost14a}. 
Dominant contributions to the final states of the strongest dipole-allowed CT excitations from the optical exciton localized on pentacene 
are shown in Figs.~3(a) and (b). Supplementary Table II gives the normalized contributions to the final states of all wavefunctions to which ESAs from the optical 
singlets are expected, and the dipole couplings between the initial and final eigenstates \cite{SM}.
Once again, CT between the terminal {\bf P} molecules is significantly smaller than that between {\bf P} and {\bf T}. Taken together, Figs.~2 and 3
give clear evidence that $^1(T_{1[P]}T_{1[P]})$ is not generated in a single-step CT-mediated process.

We have 
calculated near-exact energies and wavefunctions of the lowest triplet-triplet eigenstates of {\bf P-T-P}.
The lowest of these, $^1$(T$_{1[P]}$T$_{1[P]}$), occurs energetically
below the optical exciton, while doubly degenerate higher energy $^1$(T$_{1[P]}$T$_{1[T]}$) and $^1$(T$_{1[T]}$T$_{1[P]}$) 
states occur immediately above the optical exciton (2.4 eV versus 2.3 eV).
The wavefunctions are shown in Figs~4(a)-(c). Triplet excitations in $^1$(T$_{1[P]}$T$_{1[P]}$) are localized 
entirely on the terminal pentacenes. Higher energy triplet-triplet eigenstates consist entirely of even and odd linear combinations of $^1$(T$_{1[P]}$T$_{1[T]}$) and 
$^1$(T$_{1[T]}$T$_{1[P]}$).
Fig.~4 shows only the largest contributions to the multiexciton eigenstates. We have examined upto 15 higher order
quadruple excitations in each case, with normalized coefficients down to 0.03. Even these higher order excitations are strictly localized
either on nearest neighbor monomers or on distant monomers, but not both in the same eigenstate.  
It is now interesting to go back to Fig.~3: the admixing between the first and last terms in the two wavefunctions here give indirect 
proof of the hypothesis that initial triple-triplet multiexciton generation occurs via virtual excitation of the CT eigenstate
dipole-coupled to the optical singlet \cite{Smith13a,Berkelbach13b,Yost14a}.}

Free triplet generation from the bound triplet-triplet is preceded by spin mixing between overall spin singlet ($m=1$), 
triplet ($m=3$) and quintet ($m=5$) multiexcitons \cite{Tayebjee17a,Chen19a}. 
In Table I we have given the calculated $m=1$ and $m=5$ energies of 
$^m$(T$_{1[P]}$T$_{1[P]}$) and $^m$(T$_{1[P]}$T$_{1[T]}$). 
$\Delta_s$ between the quintet and singlet spin states in the contiguous triplet-pair $^m$(T$_{1[P]}$T$_{1[T]}$)
is large compared to room temperature, explaining the
downhill triplet migration from this state over spin dephasing \cite{Pun19a}.
Fast thermally-induced mixing between the spin states is however expected in $^m$(T$_{1[P]}$T$_{1[P]})$. 
The larger distance between the triplet excitons occupying the terminal pentacenes in 
{\bf n} $>1$ {\bf P-Tn-P} would 
lead to even smaller $\Delta_s$, explaining the very long lifetimes of the triplet-triplet in these \cite{Pun19a}.

\begin{table}[h]
\caption{Calculated energies (in eV) of $^m$(T$_{1[P]}$T$_{1[T]}$) and $^m$(T$_{1[P]}$T$_{1[P]}$) for {\bf P-T-P}, $m=$ 1 and 5, for {\it anti-} and {\it syn-} 
connectivities. The calculated energy of the pentacene singlet exciton is 2.3 eV. $\Delta_s$ is the calculated spin gap (in eV) between $m=5$ and $m=1$.} 
\centering

\begin{tabular}{|c | c c c c|}
\hline connectivity &  multiexciton & $E(m=1)$ & $E(m=5)$ & $\Delta_s$ \\[4pt] \hline
anti & $^m(T_{[1P]}T_{[1T]})$ & 2.42 & 2.46 & 0.040 \\
 & $^m(T_{[1P]}T_{[1P]})$ & 2.04 & 2.05 & 0.005 \\ \hline
syn & $^m(T_{[1P]}T_{[1T]})$ & 2.44 & 2.47 & 0.040 \\
 & $^m(T_{[1P]}T_{[1P]})$ & 2.04 & 2.04 & 0.0003 \\ \hline

\end{tabular}
\end{table}

In summary, there indeed occur distinct triplet-triplet multiexcitons in {\bf P-T-P} (and by implication, in {\bf P-Tn-P})
with triplets occurring on neigboring versus distant acene monomers. This result is reminiscent of the
classic work by Tavan and Schulten, who showed that in linear $\pi$-conjugated polymers there occurs a ``band'' of 
covalent two-photon states \cite{Tavan87a}. The use 
of the exciton basis \cite{Khan17b,Khan18a,Parenti22a} allows us to physically locate the individual triplets in the multiexciton states.
The mechanism of the lowest energy $^1$(T$_1$T$_1$) generation in iSF chromophores is indeed dependent
on the length of the bridge acene. Direct generation of the $^1$(T$_1$T$_1$) from the optical excitation has been 
suggested in the absence of a linker \cite{Sanders16b}, but this is contentious \cite{Berkelbach13b,Yost14a}. With short acenes 
(benzene, napththalene and anthracene) as linkers, bridge monomer excitons have relatively high energies. Consequently contiguous triplet-triplets, when they exist 
(see \cite{SM}, Fig S.5) occur significantly above the lowest optical exciton on the terminal chromophore and are irrelevant to iSF. 
CT states dominated by CT between terminal chromophore and bridge molecule are also high in energy. Direct CT between the terminal chromophores 
is both energetically and optically accessible in these cases and iSF is mediated by end-to-end CT, with quantum interference 
due to electron correlation playing a strong role \cite{Parenti22a}.  
With increasing length of the bridge molecule there occurs a reversal in the energy orderings of the contiguous triplet-triplet 
and the lowest CT state, a correlation effect also reminiscent of the length dependence of energy orderings of excited states
in the shortest linear polyenes \cite{Baeriswyl92a,Soos94a,Ramasesha00a,Barford05a,Hudson82a,Tavan87a}. The iSF process occurs in two steps now. First there occurs nearest neighbor CT-mediated 
generation of contiguous triplet-triplet (see Fig.~3), followed by triplet migration to the lowest energy triplet-triplet. Large (small)
$\Delta_s$ in T$_{1[P]}$T$_{1[T]}$ (T$_{1[P]}$T$_{1[P]}$) implies rapid dephasing only following the generation of T$_{1[P]}$T$_{1[P]}$. Most importantly, 
which particular mechanism dominates can be anticipated already from ground state absorption spectra: 
CT absorption below or at the molecular absorption energy due to the bridge molecule indicates CT-mediated SF that will exhibit strong
quantum interference effects. CT absorption higher in energy than the bridge exciton 
is due to CT among nearest monomers; there is no dependence of absorption strength on connectivity here, and the final state of iSF is
reached by ultrafast triplet migration. While the present theoretical work is based on acene-based compounds,
the fundamental conclusions as well as the computational approach can both be extended to other systems, thereby
contributing to the the design of new iSF chromophores.

Work at Arizona was partially supported by National Science Foundation (NSF)
grant NSF-CHE-1764152.  Some of the calculations were performed using
high performance computing resources maintained by the University of
Arizona Research Technologies department and supported by the University of
Arizona Technology and Research Initiative Fund (TRIF), University
Information Technology Services (UITS), and Research, Innovation, and
Impact (RII). P. B. acknowledges financial support from the German Research Foundation (Deutsche Forschungsgemeinschaft, DFG), Project ID 441216021, 
and U. Nitzsche for technical support.

\begin{thebibliography}{10}

\bibitem{Baeriswyl92a}
D.~Baeriswyl, D.~K. Campbell, and S.~Mazumdar.
\newblock An overview of the theory of \protect{$\pi$-Conjugated} polymers.
\newblock In H.~Kiess, editor, {\em Conjugated Conducting Polymers}. Springer
  Verlag, Berlin, 1992.

\bibitem{Soos94a}
Z.~G. Soos, D.~S. Galvao, and S.~Etemad.
\newblock Fluorescence and excited-state structure of conjugated polymers.
\newblock {\em Adv. Mater.}, 6:280--287, 1994.

\bibitem{Ramasesha00a}
S.~Ramasesha, Swapan~K. Pati, Z.~Shuai, and J.L. Br\'edas.
\newblock The density matrix renormalization group method: Application to the
  low-lying electronic states in conjugated polymers.
\newblock {\em Adv. Quant. Chem.}, pages 121 -- 215, 2000.

\bibitem{Barford05a}
W.~Barford.
\newblock {\em Electronic and Optical Properties of Conjugated Polymers}.
\newblock Oxford Science Publications, 2005.

\bibitem{Hudson82a}
B.~S. Hudson, B.~E. Kohler, and K.~Schulten.
\newblock Linear polyene electronic$-$structure and potential surfaces.
\newblock {\em Excited States}, 6:1--95, 1982.

\bibitem{Raghu02a}
C.~Raghu, Y.~Anusooya Pati, and S.~Ramasesha.
\newblock Density-matrix renormalization-group study of low-lying excitations
  of polyacene within a {P}ariser-{P}arr-{P}ople model.
\newblock {\em Phys.\ Rev.\ B}, 66:035116, 2002.

\bibitem{Sanders20a}
S.~N. Sanders, E.~Kumarasamy, K.~J. Fallon, M.~Y. Sfeir, and L.~M. Campos.
\newblock Singlet fission in a hexacene dimer: energetics dictate dynamics.
\newblock {\em Chem. Sci}, 11:1079--1084, 2020.

\bibitem{Smith13a}
M.~B. Smith and J.~Michl.
\newblock Recent advances in singlet fission.
\newblock {\em Annu.\ Rev.\ Phys.\ Chem.}, 64:361--386, 2013.

\bibitem{Lee13a}
J.~Lee, P.~Jadhav, P.~D. Reusswig, S.~R. Yost, N.~J. Thompson, D.~N. Congreve,
  E.~Hontz, T.~Van Voorhis, and M.~A. Baldo.
\newblock Singlet exciton fission photovoltaics.
\newblock {\em Acc. Chem. Res.}, 46:1300--1311, 2013.

\bibitem{Rao17a}
A.~Rao and R.~H. Friend.
\newblock Harnessing singlet exciton fission to break the
  \protect{Shockley-Queisser} limit.
\newblock {\em Nature Reviews}, 2:17063, 2017.

\bibitem{Xia17a}
J.~Xia, S.~N. Sanders, W.~Cheng, J.~Z. Low, J.~Liu, L.~M. Campos, and T.~Sun.
\newblock Singlet fission: Progress and prospects in solar cells.
\newblock {\em Adv. Mater.}, 29:1601652, 2017.

\bibitem{Casanova18a}
D.~Casanova.
\newblock Theoretical modeling of singlet fission.
\newblock {\em Chem. Rev.}, 118:7164--7207, 2018.

\bibitem{Felter19a}
K.~M. Felter and F.~C. Grozema.
\newblock Singlet fission in crystalline organic materials: Recent insights and
  future directions.
\newblock {\em J. Phys. Chem. Lett.}, 10:7208--7214, 2019.

\bibitem{Shockley61a}
W.~Shockley and H.~J. Queisser.
\newblock Detailed balance limit of efficiency of p-n junction solar cells.
\newblock {\em J. Appl. Phys.}, 32:510--519, 1961.

\bibitem{Sanders15a}
S.~N. Sanders, E.~Kumarasamy, A.~B. Pun, M.~T. Trinh, B.~Choi, J.~Xia, E.~J.
  Taffet, J.~Z. Low, J.~R. Miller, X.~Roy, X.-Y. Zhu, M.~L. Steigerwald, M.~Y.
  Sfeir, and L.~M. Campos.
\newblock Quantitative intramolecular singlet fission in bipentacenes.
\newblock {\em J.\ Am.\ Chem.\ Soc.}, 137(28):8965--8972, 2015.

\bibitem{Zirzlmeier15a}
J.~Zirzlmeier, D.~Lehnherr, P.~B. Coto, E.~T. Chernick, R.~Casillas B.~S.
  Basel, M.~Thoss, R.~R. Tykwinski, and D.~M. Guldi.
\newblock Singlet fission in pentacene dimers.
\newblock {\em Proc. Natl. Acad. Sci.}, 112(17):5325--5330, 2015.

\bibitem{Lukman15a}
S.~Lukman, A.~J. Musser, K.~Chen, A.~Stavros, C.~K. Yang, Z.~Zeng, Q.~Ye,
  C.~Chi, J.~M. Hodgkiss, J.~Wu J., R.~H. Friend, and N.~C. Greenham.
\newblock Tuneable singlet exciton fission and triplet-triplet annihilation in
  an orthogonal pentacene dimer.
\newblock {\em Adv. Funct. Mater.}, 25:5452--5461, 2015.

\bibitem{Sakuma16a}
T.~Sakuma, H.~Sakai, Y.~Araki, T.~Mori, T.~Wada, N.~Tkachenko, and T.~Hasobe.
\newblock Long-lived triplet excited states of bent-shaped pentacene dimers by
  intramolecular singlet fission.
\newblock {\em J. Phys. Chem. A}, 120:1867--1875, 2016.

\bibitem{Sanders16a}
S.~N. Sanders, E.~Kumarasamy, A.~B. Pun, M.~L. Steigerwald, M.~Y. Sfeir, and
  L.~M. Campos.
\newblock Intramolecular singlet fission in oligoacene heterodimers.
\newblock {\em Angew. Chem. Int. Ed.}, 55:3373--3377, 2016.

\bibitem{Korovina16a}
N.~V. Korovina, S.~Das, Z.~Nett, X.~Feng, J.~Joy, R.~Haiges, A.~I. Krylov,
  S.~E. Bradforth, and M.~E. Thompson.
\newblock Singlet fission in a covalently linked cofacial alkynyltetracene
  dimer.
\newblock {\em J.\ Am.\ Chem.\ Soc.}, 138:617--627, 2016.

\bibitem{Korovina18a}
N.~V. Korovina, J.~Joy, X.~T. Feng, C.~Feltenberger, A.~I. Krylov, S.~E.
  Bradforth, and M.~E. Thompson.
\newblock Linker-dependent singlet fission in tetracene dimers.
\newblock {\em J. Am. Chem. Soc.}, 140:10179--10190, 2018.

\bibitem{Zirzlmeier16a}
J.~Zirzlmeier, R.~Casillas, S.~R. Reddy, P.~B. Coto, D.~Lehnherr, E.~T.
  Chernick, I.~Papadopoulos, M.~Thoss, R.~R. Tykwinski, and D.~M. Guldi.
\newblock Solution-based intramolecular singlet fission in cross-conjugated
  pentacene dimers.
\newblock {\em Nanoscale}, 8:101133, 2016.

\bibitem{Sun16a}
T.~Sun, L.~Shen, H.~Liu, X.~Sun, and X.~Li.
\newblock Synthesis and photophysical properties of a single bond linked
  tetracene dimer.
\newblock {\em J. Mol. Struc.}, 1116:200--206, 2016.

\bibitem{Basel17a}
Bettina~S. Basel, Johannes Zirzlmeier, Constantin Hetzer, Brian~T. Phelan,
  Matthew~D. Krzyaniak, S.~Rajagopala Reddy, Pedro~B. Coto, Noah~E. Horwitz,
  Ryan~M. Young, Fraser~J. White, Frank Hampel, Timothy Clark, Michael Thoss,
  Rik~R. Tykwinski, Michael~R. Wasielewski, and Dirk~M. Guldi.
\newblock Unified model for singlet fission within a non-conjugated covalent
  pentacene dimer.
\newblock {\em Nat.\ Commun.}, 8:15171, 2017.

\bibitem{Margulies16a}
E.~A. Margulies, C.~E. Miller, Y.~Wu, L.~Ma, G.~C. Schatz, R.~M. Young, and
  M.~R. Wasielewski.
\newblock Enabling singlet fission by controlling intramolecular $\pi$ stacked
  covalent terrylenediimide dimers.
\newblock {\em Nat.\ Chem.}, 8:1120--1125, 2016.

\bibitem{Pun19a}
A.~B. Pun, A.~Asadpoordarvish, E.~Kumarasamy, M.~J.~Y. Tayebjee, D.~Niesner,
  D.~R. McCamey, S.~N. Sanders, L.~M. Campos, and M.~Y. Sfeir.
\newblock Ultra-fast intramolecular singlet fission to persistent multiexcitons
  by molecular design.
\newblock {\em Nat.\ Chem.}, 11:821--828, 2019.

\bibitem{Krishnapriya19a}
K.C. Krishnapriya, Palas Roy, Boregowda Puttaraju, Ulrike Salzner, Andrew~J.
  Musser, Manish Jain, Jyotishman Dasgupta, and Satish Patil.
\newblock Spin density encodes intramolecular singlet exciton fission in
  pentacene dimers.
\newblock {\em Nat.\ Commun.}, 10:33, 2019.

\bibitem{Korovina18c}
N.~V. Korovina, Pompetti~N. F., and Johnson~J. C.
\newblock Lessons from intramolecular singlet fission with covalently bound
  chromophores.
\newblock {\em J.\ Chem.\ Phys.}, 152:040904, 2020.

\bibitem{Hetzer18a}
C.~Hetzer, D.~M. Guldi, and R.~R. Tykwinski.
\newblock Pentacene dimers as a critical tool for the investigation of
  intramolecular singlet fission.
\newblock {\em Chem. Eur. J.}, 24:8245--8257, 2018.

\bibitem{Chen18a}
M.~Chen, Y.~J. Bae, C.~M. Mauck, A.~Mandal, R.~M. Young, and M.~R. Wasielewski.
\newblock Singlet fission in covalent terrylenediimide dimers: Probing the
  nature of the multiexciton state using femtosecond mid-infrared spectroscopy.
\newblock {\em J. Am. Chem. Soc.}, 140:9184--9192, 2018.

\bibitem{Parenti20a}
K.~R. Parenti, G.~He, S.~N. Sanders, A.~B. Pun, E.~Kumarasamy, M.~Y. Sfeir, and
  L.~M. Campos.
\newblock Bridge resonance effects in singlet fission.
\newblock {\em J. Phys. Chem. A}, 124:9392--9399, 2020.

\bibitem{Wang22a}
Kangwei Wang, Guangwei Shao, Shaoqian Peng, Xiaoxiao You, Xingyu Chen, Jingwen
  Xu, Huaxi Huang, Huan wang, Di~Wu, and Jianlong Xia.
\newblock Achieving symmetry-braking charge separation in perylenediimide
  trimers: The effect of bridge resonance.
\newblock {\em J. Phys. Chem. B}, 126:3758--3767, 2022.

\bibitem{Purdy23a}
M.~Purdy, P.~Budden, K.~Fallon, Cara~N. Gannett, H.~D. Abruña, W.~Zeng,
  R.~Friend, A.~J. Musser, and H.~Bronstein.
\newblock Re-thinking dimer design principles with indolonaphthyridine
  intramolecular singlet fission.
\newblock {\em Chem. Eur. J.}, page e202301547, 2023.

\bibitem{Majumder23a}
K.~Mazjumder, S.~Mukherjee, N.~A. Panjwani, J.~Lee, R.~Bittl, W.~Kim, S.~Patil,
  and A.~J. Musser.
\newblock Controlling intramolecular singlet fission dynamics via torsional
  modulation of through-bond versus through-space couplings.
\newblock {\em J.\ Am.\ Chem.\ Soc.}, 145:20883--208896, 2023.

\bibitem{Khan17b}
S.~Khan and S.~Mazumdar.
\newblock Diagrammatic exciton basis theory of the photophysics of pentacene
  dimers.
\newblock {\em J. Phys. Chem. Lett.}, 8:4468--4478, 2017.

\bibitem{Khan18a}
S.~Khan and S.~Mazumdar.
\newblock Optical probes of the quantum-entangled triplet-triplet state in a
  heteroacene dimer.
\newblock {\em Phys.\ Rev.\ B}, 98:165202, 2018.

\bibitem{Trinh17a}
M.~T. Trinh, A.~Pinkard, A.~B. Pun, S.~N. Sanders, E.~Kumarasamy, M.~Y. Sfeir,
  L.~M. Campos, X.~Roy, and X.-Y. Zhu.
\newblock Distinct properties of the triplet pair state from singlet fission.
\newblock {\em Science Advances}, 3(7):e1700241, 2017.

\bibitem{Miyata19a}
K.~Miyata, F.~S. Conrad-Burton, F.~L. Geyer, and X.-Y. Zhu.
\newblock Triplet pair states in singlet fission.
\newblock {\em Chem. Rev.}, 119:4261--4292, 2019.

\bibitem{Pensack18a}
Ryan~D. Pensack, Andrew~J. Tilley, Christopher Grieco, Geoffrey~E. Purdum,
  Evgeny~E. Ostroumov, Devin~B. Granger, Daniel~G. Oblinsky, Jacob~C. Dean,
  Grayson~S. Doucette, John~B. Asbury, Yueh-Lin Loo, Dwight~S. Seferos, John~E.
  Anthony, and Gregory~D. Scholes.
\newblock Striking the right balance of intermolecular coupling for
  high-efficiency singlet fission.
\newblock {\em Chem. Sci.}, 9:6240--6259, 2018.

\bibitem{Masoomi-Godarzi20a}
S.~Masoomi-Godarzi, C.~R. Hall, B.~Zhang, M.~A. Gregory, J.~M. White, W.~W.~H.
  Wong, K.~P. Ghiggino, T.~A. Smith, and D.~A. Jones.
\newblock Competitive triplet formation and recombination in crystalline films
  of perylenediimide derivatives: Implications for singlet fission.
\newblock {\em J. Phys. Chem. C}, 124:11574, 2020.

\bibitem{Kim18a}
H.~Kim and P.~M. Zimmerman.
\newblock Coupled double triplet state in singlet fission.
\newblock {\em Physical Chemistry Chemical Physics}, 20:30083--30094, 2018.

\bibitem{Musser19a}
A.~J. Musser and J.~Clark.
\newblock Triplet-pair states in organic semiconductors.
\newblock {\em Annu. Rev. Phys. Chem.}, 70:323–351, 2019.

\bibitem{Scholes15a}
G.~D. Scholes.
\newblock Correlated pair states formed by singlet fission and exciton-exciton
  annihilation.
\newblock {\em J. Phys. Chem. A}, 119:12699, 2015.

\bibitem{Pensack16a}
R.~D. Pensack, E.~E. Ostroumov, A.~J. Tilley, S.~Mazza, C.~Grieco, K.~J.
  Thorley, J.~B. Asbury, D.~S. Seferos, J.~E. Anthony, and G.~D. Scholes.
\newblock Observation of two triplet-pair intermediates in singlet exciton
  fission.
\newblock {\em J. Phys. Chem. Lett.}, 7:2370--2375, 2016.

\bibitem{Lee18a}
T.~S. Lee, Y.~L. Lin, H.~Kim, R.~D. Pensack, B.~P. Rand, and G.~D. Scholes.
\newblock Triplet energy transfer governs the dissociation of the correlated
  triplet pair in exothermic singlet fission.
\newblock {\em J. Phys. Chem. Lett.}, 9:4087--4095, 2018.

\bibitem{Taffet20a}
E.~J. Taffet, D.~Beljonne, and G.D. Scholes.
\newblock Overlap-driven splitting of triplet pairs in singlet fission.
\newblock {\em J.\ Am.\ Chem.\ Soc.}, 142:20040--20047, 2020.

\bibitem{Hudson22a}
R.~J. Hudson, A.~N. Stuart, D.~M. Huang, and T.~W. Kee.
\newblock What next for singlet fission in photovoltaics?
\newblock {\em J. Phys. Chem. C}, 126:5369--5377, 2022.

\bibitem{Parenti22a}
K.~Parenti, R.~Chesler, G.~He, P.~Bhattacharyya, B.~Xiao, D.~Malinowski,
  J.~Zhang, X.~Yin, A.~Shukla, S.~Mazumdar, M.~Sfeir, and L.~Campos.
\newblock The role of quantum interference in intramolecular singlet fission.
\newblock {\em Nat.\ Chem.}, 15:339--346, 2022.

\bibitem{Pariser53a}
R.~Pariser and R.G. Parr.
\newblock A semi-empirical theory of the electronic spectra and electronic
  structure of complex unsaturated molecules ii.
\newblock {\em J.\ Chem.\ Phys.}, 21:767--776, 1953.

\bibitem{Pople53a}
J.~A. Pople.
\newblock Electron interaction in unsaturated hydrocarbons.
\newblock {\em Trans. Faraday Soc.}, 49:1375--1385, 1953.

\bibitem{Ramasesha90a}
S.~Ramasesha and I.D.L. Albert.
\newblock Sudden polarization in interacting model $\pi$-systems: An exact
  study.
\newblock {\em Chem.\ Phys.}, 142(3):395 -- 402, 1990.

\bibitem{Chandross97a}
M.~Chandross and S.~Mazumdar.
\newblock {C}oulomb interactions and linear, nonlinear, and triplet absorption
  in poly(para-phenylenevinylene).
\newblock {\em Phys.\ Rev.\ B}, 55:1497--1504, 1997.

\bibitem{Tavan87a}
P.~Tavan and K.~Schulten.
\newblock Electronic excitations in finite and infinite polyenes.
\newblock {\em Phys.\ Rev.\ B}, 36:4337--4358, 1987.

\bibitem{Aryanpour15a}
Karan Aryanpour, Alok Shukla, and Sumit Mazumdar.
\newblock Theory of singlet fission in polyenes, acene crystals, and covalently
  linked acene dimers.
\newblock {\em J. Phys. Chem. C}, 119(13):6966--6979, 2015.

\bibitem{SM}
See Supplemental Material at http://link.aps.org/ supplemental/xx.xxxx/
  PhysRevLett.xxx.xxxxxx for further details of calculations and discussion of
  relevance to experiments.

\bibitem{Berkelbach13b}
Timothy~C. Berkelbach, Mark~S. Hybertsen, and David~R. Reichman.
\newblock Microscopic theory of singlet exciton fission. ii. application to
  pentacene dimers and the role of superexchange.
\newblock {\em J.\ Chem.\ Phys.}, 138(11):114103, 2013.

\bibitem{Yost14a}
S.~R. Yost, J.~Lee, M.~W.~B. Wilson, T.~Wu, D.~P. McMahon, R.~R. Parkhurst,
  Nicholas~J. Thompson, Daniel~N. Congreve, A.~Rao, K.~Johnson, M.~Y. Sfeir,
  M.~G. Bawendi, T.~M. Swager, R.~H. Friend, M.~A. Baldo, and T.~Van~Voorhis.
\newblock A transferable model for singlet-fission kinetics.
\newblock {\em Nat. Chem.}, 6:492--497, 2014.

\bibitem{Tayebjee17a}
Murad J.~Y. Tayebjee, S.~N. Sanders, E.~Kumaraswamy, L.~M. Campos, M.~Y. Sfeir,
  and D.~R. McCamey.
\newblock Quintet multiexciton dynamics in singlet fission.
\newblock {\em Nat. Phys.}, 13:182--188, 2017.

\bibitem{Chen19a}
M.~Chen, M.~D. Krzyaniak, J.~N. Nelson, Y.~J. Bae, S.~M. Harveya, R.~D.
  Schaller, R.~M. Young, and M.~R. Wasielewski.
\newblock Quintet-triplet mixing determines the fate of the multiexciton state
  produced by singlet fission in a terrylenediimide dimer at room temperature.
\newblock {\em Proc. Natl. Acad. Sci. USA}, 116:8178–8183, 2019.

\bibitem{Sanders16b}
E.~G. Fuemmeler, S.~N. Sanders, A.~B. Pun, E.~Kumarasamy, T.~Zeng, K.~Miyata,
  M.~L. Steigerwald, X.~Y. Zhu, M.~Y. Sfeir, L.~M. Campos, and N.~Ananth.
\newblock A direct mechanism of ultrafast intramolecular singlet fission in
  pentacene dimers.
\newblock {\em ACS Cent. Sci.}, 2(5):316--324, 2016.

\end{thebibliography}

\end{document}